\documentclass[twocolumn]{aastex6}
\pdfoutput=1 
\usepackage{amsmath,amstext}
\usepackage{apjfonts} 
\usepackage{afterpage}

\usepackage{xcolor}

\shorttitle{Late-Time Behavior of GW\,170817}
\shortauthors{Alexander et al.}

\begin{document}

\title{A Decline in the X-ray through Radio Emission from GW170817 Continues to Support an Off-Axis Structured Jet}

\author{K.~D.~Alexander\altaffilmark{1}, {R.~Margutti}\altaffilmark{2}, {P.~K.~Blanchard}\altaffilmark{1}, W.~Fong\altaffilmark{2,3},  E.~Berger\altaffilmark{1}, {A.~Hajela}\altaffilmark{2}, {T.~Eftekhari}\altaffilmark{1}, {R.~Chornock}\altaffilmark{4}, {P.~S.~Cowperthwaite}\altaffilmark{1},  {D.~Giannios}\altaffilmark{5}, {C.~Guidorzi}\altaffilmark{6}, {A.~Kathirgamaraju}\altaffilmark{5}, {A.~MacFadyen}\altaffilmark{7}, {B.~D.~Metzger}\altaffilmark{8}, {M.~Nicholl}\altaffilmark{1}, {L.~Sironi}\altaffilmark{9},  {V.~A.~Villar}\altaffilmark{1}, {P.~K.~G.~Williams}\altaffilmark{1}, {X.~Xie}\altaffilmark{7}, {J.~Zrake}\altaffilmark{8}}

\altaffiltext{1}{Harvard-Smithsonian Center for Astrophysics, 60 Garden Street, Cambridge, Massachusetts 02138, USA}
\altaffiltext{2}{Center for Interdisciplinary Exploration and Research in Astrophysics (CIERA) and Department of Physics and Astronomy, Northwestern University, Evanston, IL 60208}
\altaffiltext{3}{Hubble Fellow}
\altaffiltext{4}{Astrophysical Institute, Department of Physics and Astronomy, 251B Clippinger Lab, Ohio University, Athens, OH 45701, USA}
\altaffiltext{5}{Department of Physics and Astronomy, Purdue University, 525 Northwester Avenue, West Lafayette, IN 47907, USA}
\altaffiltext{6}{Department of Physics and Earth Science, University of Ferrara, via Saragat 1, I--44122, Ferrara, Italy}
\altaffiltext{7}{Center for Cosmology and Particle Physics, New York University, 726 Broadway, New York, NY 10003, USA}
\altaffiltext{8}{Department of Physics and Columbia Astrophysics Laboratory, Columbia University, New York, NY 10027, USA}
\altaffiltext{9}{Columbia University, Pupin Hall, 550 West 120th Street, New York, NY 10027, USA}

\begin{abstract}
We present new observations of the binary neutron star merger GW170817 at $\Delta t\approx 220-290$ days post-merger, at radio (Karl G.~Jansky Very Large Array; VLA), X-ray ({\it Chandra X-ray Observatory}) and optical ({\it Hubble Space Telescope}; HST) wavelengths.  These observations provide the first evidence for a turnover in the X-ray light curve, mirroring a decline in the radio emission at $\gtrsim5\sigma$ significance. The radio-to-X-ray spectral energy distribution exhibits no evolution into the declining phase. Our full multi-wavelength dataset is consistent with the predicted behavior of our previously published models of a successful structured jet expanding into a low-density circumbinary medium, but pure cocoon models with a choked jet cannot be ruled out. If future observations continue to track our predictions, we expect that the radio and X-ray emission will remain detectable until $\sim 1000$ days post-merger.
\end{abstract}


\section{Introduction}

The discovery of broadband synchrotron emission associated with the binary neutron star merger, GW170817 \citep{lvc170817,em170817}, together with the prompt low-luminosity gamma-ray emission \citep{fermi170817} provided the first direct evidence for the production of relativistic ejecta in such a system \citep{alex17,ggg17,Haggard17,hal17,Kasliwal17,marg17,marg18,mur17,Troja17,kat18,mool18}. Several lines of evidence suggest that short gamma-ray bursts (SGRBs) are produced by binary neutron star mergers viewed on-axis \citep{ber14}, and observations of SGRB afterglows have provided a measure of the kinetic energies, collimation angles, and circumbinary densities \citep{fong15}.  These observations provide a basis for comparison with GW170817, suggesting that viewing angle effects are the dominant cause of observed differences between GW170817 and these events (e.g., \citealt{fong17}). 

In GW170817 the radio and X-ray emission were observed to rise gradually for the first $\sim 160$ days, with a single spectral power law, $f_{\nu}\propto \nu^{\beta}$ with $\beta\approx -0.6$, spanning $\sim 10^9-10^{18}$ Hz \citep{alex17,hal17,kim17,dav18,marg18,mool18,Pooley17,res18,Ruan18,Troja18}. This emission was accompanied by optical detections well after the radioactive ``kilonova'' component had faded away \citep{lyman18,marg18}.  These observed properties can be explained in the context of two primary models.  First is a ``successful structured jet'' with properties similar to those inferred in SGRBs but viewed at an angle $\theta_{\rm obs}\approx 20^\circ - 30^\circ$ off-axis \citep{lk17,laz17,marg18,xie18}. Successful structured jets consist of an initially highly relativistic, collimated core surrounded by wings of mildly relativistic material at larger angles (e.g. \citealt{ros02,kg03}); they are also sometimes called ``successful jets with cocoons" in the literature {(e.g. \citealt{np18})}. The alternative is that no successful jet was produced and the emission is quasi-isotropic. 

\clearpage
\begin{longtable*}{cccccccc}
\caption{New Observations of GW170817} 
\label{tab:obs} \\
\hline
Observatory & 
UT Date & 
$\Delta t$ & 
Mean Freq. & 
Freq.~Range/ & 
Exp.~Time  & 
Flux Density &{Image RMS}\\        
 & 
 (UT)  & 
 (days) & 
 (Hz)   &  
 Filter & 
 (hr) & 
  ($\mu$Jy) &  
 {($\mu$Jy)} \\
\hline
VLA & 2018 Mar 22 & 216.91 & $3.0\times10^9$	& $2-4$ GHz & 0.6 & $69\pm15$ & $10$\\
VLA & 2018 Mar 22 & 216.88 & $6.0\times10^9$	& $4-8$ GHz  & 0.7 & $39\pm9$ & $6$\\
VLA & 2018 Mar 22 & 216.85 & $10.0\times10^9$	& $8-12$ GHz & 0.7 & $28\pm7$ & $5$\\
VLA & 2018 Mar 22 & 216.80 & $15.0\times10^9$	& $12-18$ GHz & 1.8 & $21\pm5$ & $4$\\
VLA & 2018 May 1 & 256.76 & $3.0\times10^9$ & $2-4$ GHz & 0.7 & $55\pm12$ & $9$\\
VLA & 2018 May 17 & 272.67 & $3.0\times10^9$ & $2-4$ GHz & 1.3 & $44\pm11$ & $8$\\
VLA & 2018 May 17 & {272.61} & $6.0\times10^9$ & $4-8$GHz & 1.3 & ${36\pm7}$ & ${5}$\\
VLA & {2018 Jun 2} & {288.61} & ${3.0\times10^9}$ & ${2-4}$ {GHz} & {1.3} & ${46\pm11}$ & ${8}$\\
VLA & {2018 Jun 2} & {288.55} & ${6.0\times10^9}$ & ${4-8}$ {GHz} & {1.3} & ${35\pm7}$ & ${5}$\\
\hline{\it HST} & 2018 Mar 23 & 218.37 & $5.07\times10^{14}$ & F606W & 0.58 & $<0.070^{a}$  & ${0.023}$\\\hline
{\it Chandra} & 2018 May 3--5 & 259.99 & $2.4\times10^{17}$ & $0.3-10$ keV & 27.2 &  $1.22^{+0.25}_{-0.15}\times 10^{-3}$ & ${-}$ \\
\hline
\noalign{\smallskip}
\noalign{\text{$^{a}$ Corrected for Galactic extinction.}}
\end{longtable*}

\noindent In this case, either a choked jet deposits all of its energy into a mildly relativistic cocoon that dominates the emission (referred to as a ``pure cocoon'' model in this Letter; \citealt{Kasliwal17,got17,mool18}) or the emission arises from the fastest component of the dynamical ejecta expelled during the merger \citep{hot18,mool18}. We note that independent of the radio/X-ray data, a recent joint analysis of the Laser Interferometer Gravitational-Wave Observatory (LIGO)/Virgo gravitational wave data and the electromagnetic observations indicates that the inclination angle of the binary is $32^{+10}_{-13}$ deg \citep{finstad18}, in agreement with the viewing angle inferred from the {successful} structured jet model. {Recently reported very long baseline interferometry (VLBI) observations of GW170817 also support the existence of a successful jet with a viewing angle of ${20^\circ\pm5^\circ}$ \citep{vlbi}. }

A possible way to distinguish the origin of the relativistic ejecta is to measure the long-term behavior of the post-peak emission (e.g., \citealt{gg18,marg18,np18,Troja18}). Recently, a turnover in the radio light curve at $\approx 200$ days  was reported \citep{dob18}, while {\it Chandra} and {\it XMM-Newton} observations extending to $\approx 160$ days have been suggestive of a flattening in the X-ray light curve \citep{dav18,marg18}.  Here, we report new radio, optical, and X-ray observations at $\approx 220-{290}$ days that unambiguously show a decline at both X-ray and radio wavelengths, as well as a possible decline in the optical band. The declining behavior is fully consistent with the predicted behavior of our {successful} structured jet models \citep{marg18,xie18}.
{Uncertainties are $1\sigma$ confidence intervals unless otherwise specified.}

\section{Observations}

We present new radio, optical, and X-ray observations of GW170817, and analyze those jointly with all previous observations from our work and from the literature \citep{alex17,Haggard17,marg17,marg18,dav18,dob18,lyman18,mool18}.

\subsection{VLA Observations}

We observed GW170817 with the VLA on 2018 March 22 UT in the A configuration, with nearly continuous frequency coverage at $2-18$ GHz (apart from small gaps introduced by radio frequency interference (RFI) and correlator setup), on May 1 UT at $2-4$ GHz, {and on May 17 UT and June 2 UT at ${2-8}$ GHz}. We analyzed the data with CASA \citep{casa} using 3C286 as the bandpass calibrator and J$1258-2219$ as the phase calibrator.  We imaged the data using standard CASA routines, using a bandwidth of 2, 4, or 6 GHz for each image, and determined the flux density by fitting a point source model using the {\tt imtool} program within the {\tt pwkit} package \citep{pwkit}. {Our March 22 3 GHz observation was impacted by unusually strong RFI, resulting in elevated noise in the image produced after our initial data reduction. We therefore reprocessed the data using a thorough independent manual flagging procedure.} The {final} flux density values are provided in \autoref{tab:obs}. 

We also measure the flux density of the compact background source {at (R.A., decl.) ${= 13^{\rm h}9^{\rm m}53^{\rm s}.911, -23^{\circ}21\arcmin34\arcsec.49}$ (J2000)} noted by \cite{dob18} and find that it remains constant in comparison to our previous observations of this field. {For the new data presented here, we obtain ${F_{\nu}=599\pm5}$ ${\mu}$Jy at 3 GHz, ${F_{\nu}=370\pm5}$ ${\mu}$Jy at 6 GHz, ${F_{\nu}=146\pm8}$ ${\mu}$Jy at 10 GHz, and ${F_{\nu}=57\pm5}$ ${\mu}$Jy at 15 GHz, where the first two measurements are average values and the second two come from our observations at 217 days.}

\subsection{{\it HST} Observations}
We obtained one orbit of \textit{HST} observations with the Advanced Camera for Surveys (ACS) Wide Field Camera on 2018 March 23 UT using the F606W filter (PID: 15329; PI: Berger).  We analyze the data in the same manner as our 2018 January observation described in \citet{marg18}.  We do not detect a source at the position of GW170817 and determine the limiting magnitude by injecting point sources of varying luminosities at the position of GW170817 and then performing galaxy subtraction using {\tt GALFIT} v3.0.5 \citep{Peng2010} to model and remove the large-scale surface brightness profile of NGC 4993.  We measure a $3\sigma$ limit of $m_{\rm F606W}\gtrsim 27.1$ mag, calibrated to the ACS/F606W AB magnitude zeropoint provided by STScI.  After correcting for a Galactic extinction of $E(B-V$) = 0.105 mag \citep{SF2011}, this corresponds to $m_{\rm F606W} \gtrsim 26.8$ mag. Relative to our detection in 2018 January with $m_{\rm F606W}=26.60\pm 0.25$ mag \citep{marg18}, and the 2017 December detection from \citet{lyman18} with $m_{\rm F606W}=26.44\pm 0.14$ mag, the new limit is indicative of declining or flat optical brightness. 

We also subtracted the 2018 January and March images using the {\tt HOTPANTS} package \citep{hotpants}.  After performing forced aperture photometry at the position of GW170817, the residual flux in the subtracted image does not differ significantly from zero. {This is consistent with the March upper limit derived above and does not preclude a fading source}, but a definitive decline in the optical brightness relative to the January detection cannot be claimed. 

\subsection{{\it Chandra} Observations}
The {\it Chandra} X-ray Observatory (CXO) started observing GW170817 on 2018 May  03, starting at 10:41:26 UT ($t\approx 259$ days after merger) for a total exposure time of 50.8 ks (PI Wilkes; program 19408644; observation ID 21080).  \emph{Chandra} Advanced CCD Imaging Spectrometer (ACIS)-S data were reduced with the {\tt CIAO} software package (v4.9) and relative calibration files, applying standard ACIS data filtering. An X-ray source is clearly detected with {\tt wavdetect} at the location of GW170817 with significance of  $13.8\,\sigma$ and count-rate $(7.75\pm 1.28)\times 10^{-4}\,\rm{c\,s^{-1}}$  in the 0.5-8 keV energy band. A second Chandra observation was acquired on 2018 May  05, 01:25:30 UT (ID 21090, exposure time of 46.0 ks). GW170817 is detected with confidence of $14.75\,\sigma$ and 0.5-8 keV count-rate of $(8.31\pm 1.37)\times 10^{-4}\,\rm{c\,s^{-1}}$.

{For each observation we extract a spectrum using a source region of $1.5''$ and a background region of $22''$.}
We employ Cash statistics and fit the joint spectrum {with Xspec} with an absorbed power-law model with index $\Gamma$ and Galactic neutral hydrogen column density $\rm{NH}_{mw} = 0.0784\times10^{22}\,\rm{cm^{-2}}$ (\citealt{Kalberla05}).
Using Markov chain Monte Carlo (MCMC) sampling to constrain the spectral parameters we find $\Gamma=1.51^{+0.26}_{-0.27}$ 
and no statistical evidence for intrinsic neutral hydrogen absorption ($\rm{NH_{int}}<1.2 \times 10^{22}\,\rm{cm^{-2}}$ at $3\,\sigma$ c.l.).  
For these parameters, the $0.3-10$ keV  flux
is $(11.3- 15.6)\times10^{-15}\,\rm{erg\,s^{-1}cm^{-2}}$ ($1\,\sigma$ c.l.), and the unabsorbed flux is 
$(12.3- 16.9)\times10^{-15}\,\rm{erg\,s^{-1}cm^{-2}}$.  
Finally, we investigate the presence of temporal variability on short timescales and conclude that there is no evidence for statistically significant temporal variability on timescales $\ge 1$ ks.

\begin{figure}
\centerline{\includegraphics[width=0.45\textwidth]{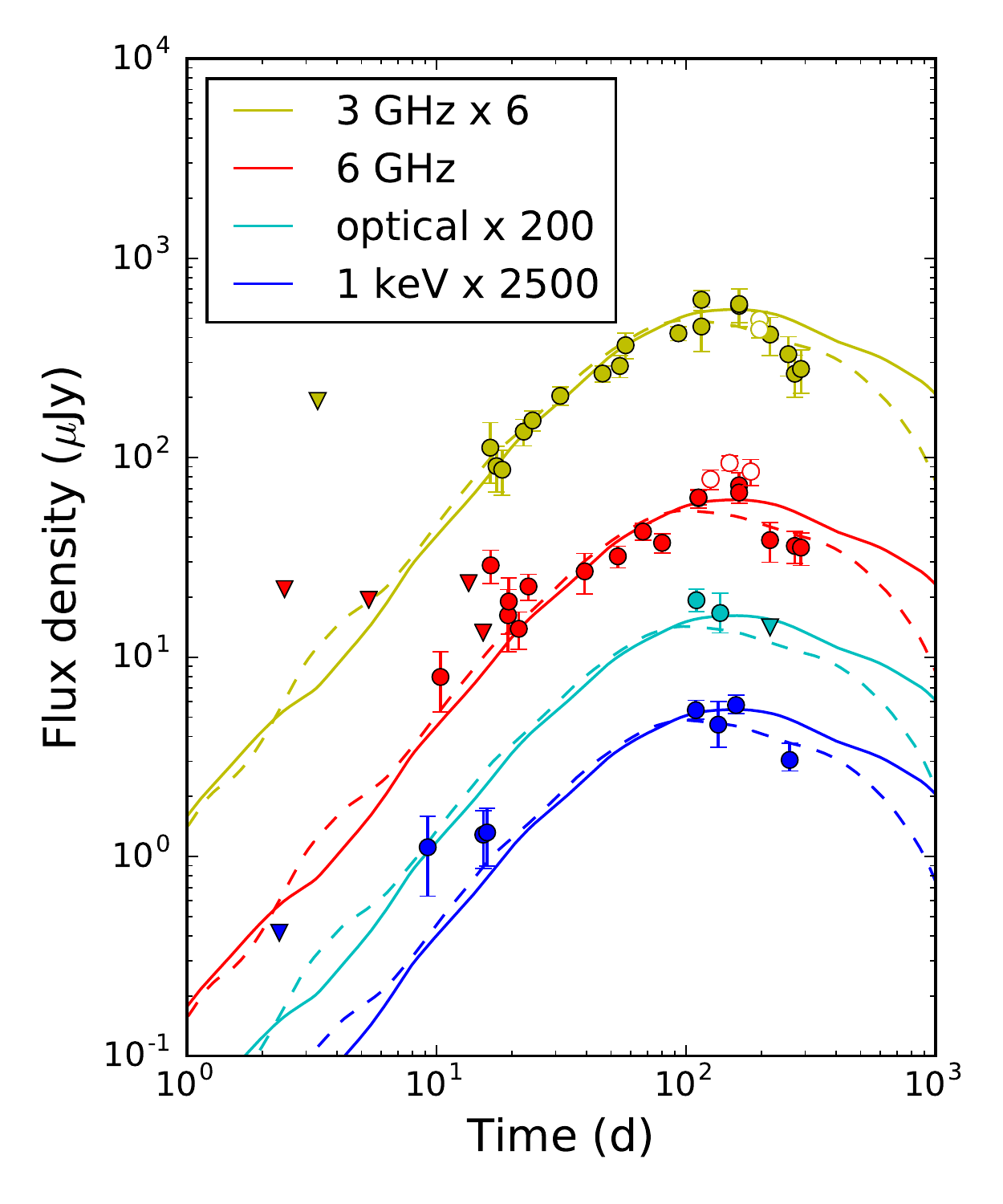}}
\caption{Up-to-date X-ray, optical, and radio light curves of GW170817 (solid circles; open circles are the new data presented in \citealt{dob18}).  The data are clearly indicative of a decline at $\gtrsim 200$ days.  Also shown are our structured jet models from \citet{marg18}; see \cite{xie18} for full details of the simulations. Both jets have an ultra-relativistic core with $E_{\rm K, iso}=6\times10^{52}$ erg within an opening angle $\theta_{\rm jet}=9^{\circ}$. The solid lines are for a model with $n=10^{-5}$ cm$^{-3}$, $\theta_{\rm obs}=17^{\circ}$, $\epsilon_e=0.1$, and $\epsilon_B=0.0005$, while the dashed lines are for $n=10^{-4}$ cm$^{-3}$, $\theta_{\rm obs}=20^{\circ}$, $\epsilon_e=0.02$, and $\epsilon_B=0.001$.  Our new radio, optical, and X-ray observations continue to support these models. }
\label{fig:jet}
\end{figure}

\section{Results and Comparison to Models}

The X-ray, optical, and 3 and 6 GHz radio light curves are shown in Figure~\ref{fig:jet}, together with our {successful} structured jet models previously presented in \cite{marg18} and \cite{xie18}. Both radio light curves show clear evidence of a decline at $\gtrsim 200$ days; the 3 GHz flux density at {289} days is about a factor of ${3}$ times fainter than its peak brightness at 163 days. To quantify the significance of this turnover, we scale all radio data {from Table \ref{tab:obs} and previous results \citep{alex17,hal17,dob18,marg18,mool18}} to a common frequency of 5.5 GHz using a spectral index of $\beta=-0.585$ \citep{marg18} and fit the resulting light curve with a smoothed broken power law 
\begin{equation}
F_\nu(t) = F_{\nu,b}\left[\frac{1}{2}\left(\frac{t}{t_b}\right)^{-s\alpha_1} + \frac{1}{2}\left(\frac{t}{t_b}\right)^{-s\alpha_2}\right]^{-1/s},
\end{equation}
where $\alpha_1$ and $\alpha_2$ are the temporal indices before and after the break time, $t_b$, respectively, $F_{\nu,b}$ is the flux density at the time of the break, and $s$ defines the sharpness of the transition. We additionally model possible calibration differences between observations taken at different facilities and reduced by different groups as an extra fractional uncertainty, $f$, on all data points\footnotemark\footnotetext[1]{We use this statistically robust approach instead of assuming a fixed fractional uncertainty of $3-5\%$ at each frequency as done by \citet{dob18}.}. We use {\tt emcee} \citep{emcee} to perform a MCMC analysis to determine the posteriors of the model parameters. We use a logarithmic flat prior on $s$ ($-1.5 < \log s < 1.5$) because by definition $s$ must be positive and the data do not provide any constraints once $s$ becomes sufficiently large (corresponding to a sharp break that occurs over a time interval much smaller than our sampling time). We use flat linear priors on all other parameters: $10<t_b<1000$ days, $1 < F_{\nu,b} < 1000$ $\mu$Jy, $-20<\alpha_1,\alpha_2 < 20$, and $0 < f < 1$. 

\begin{figure*}
\centerline{\includegraphics[width=0.95\textwidth]{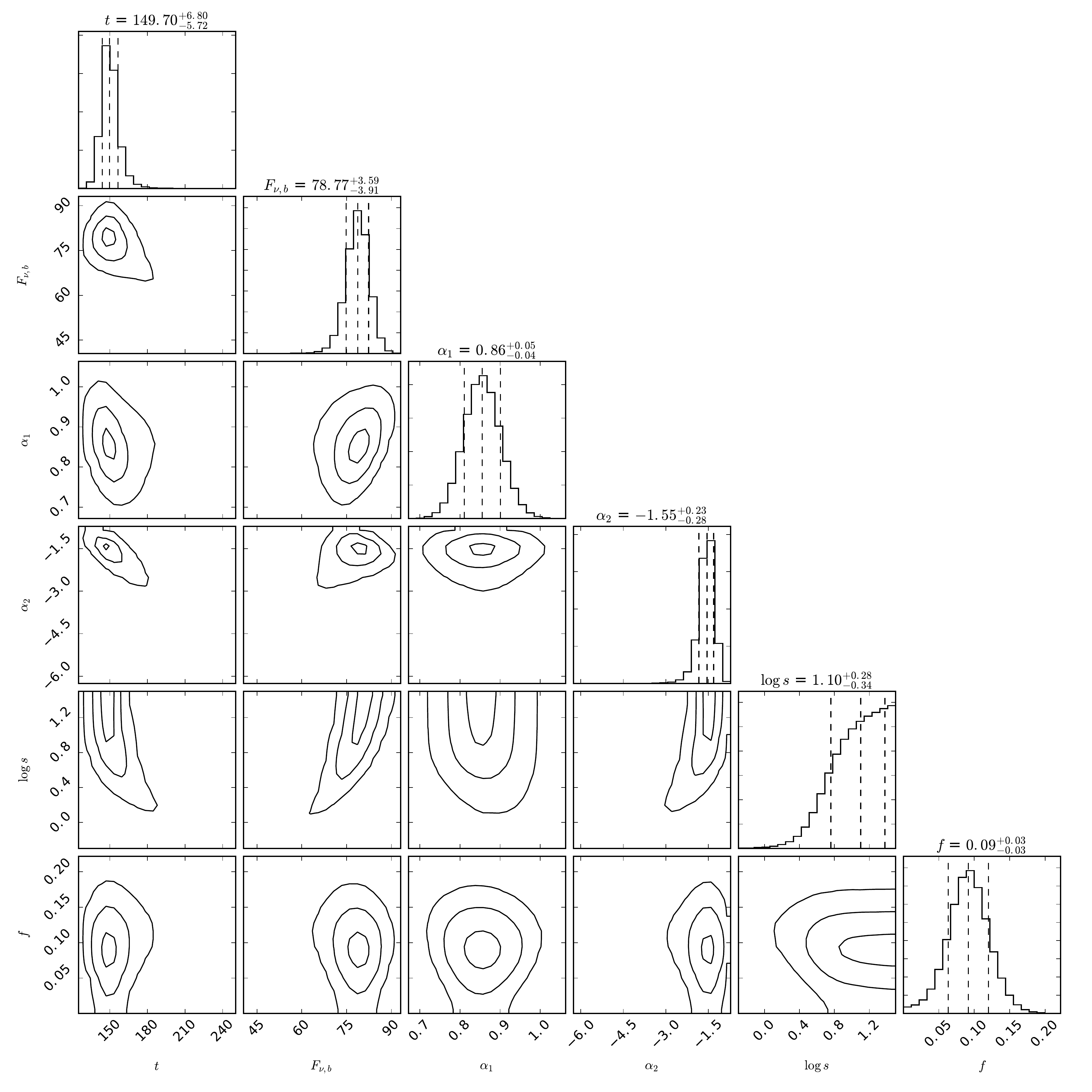}}
\caption{Posterior distributions from our MCMC analysis of the radio data.  Contours are $1$, $2$, and $3\sigma$. The data clearly prefer a sharp break, but cannot discriminate between values of $s \gtrsim10$ because this corresponds to a break much sharper than our sampling time.}
\label{fig:MCMC}
\end{figure*}

We find ${\alpha_1=0.86^{+0.05}_{-0.04}}$ and ${\alpha_2=-1.6^{+0.2}_{-0.3}}$, with ${t_b=150^{+7}_{-6}}$ days, indicative of a clear transition from a rising to a declining light curve; see Figure~\ref{fig:MCMC}.  We can reject a non-declining light curve (i.e., $\alpha_2\ge 0$) at ${7\sigma}$ confidence level.  These results are consistent\footnotemark\footnotetext[2]{We repeat our MCMC analysis for the subset of radio data presented in \citet{dob18}, with $\alpha_1$, $\alpha_2$, $t_b$, $F_{\nu,b}$, $s$, and $f$ as free parameters, and find that $t_b$ and $\alpha_2$ are more loosely constrained: $t_b=190\pm30$ days and $\alpha_2=-6^{+5}_{-8}$.  A value of $\alpha_2\ge 0$ can be ruled out at $3.0\sigma$ confidence level.} with the analysis of \citet{dob18}, but provide a much stronger indication of a break due to the longer time baseline of our observations. Both successful structured jet models and pure cocoon models are consistent with the observed ${\alpha_1}$. Unlike the simplest analytic models, which predict ${\alpha_2\approx-2.2}$ {\citep{sari99,np18}}, the \cite{xie18} successful structured jet simulations predict a shallower initial post-peak decline (Figure \ref{fig:jet}). Therefore, our derived ${\alpha_2}$ does not strongly discriminate between successful structured jet and pure cocoon or dynamical ejecta models, although the latter are mildly disfavored {\citep{lmr18}}. {Of the two \cite{xie18} models shown in Figure \ref{fig:jet}, our most recent data at 273 days and 289 days mildly favor the dashed model, which in comparison to the solid model implies a higher circumbinary density of ${10^{-4}}$ cm${^{-3}}$, a lower fraction of the shock energy imparted to electrons ${(\epsilon_e = 0.02)}$, and an observer viewing angle of ${20^\circ}$. The model light curves diverge more at later times, so future observations will allow us to place tighter constraints on these parameters.} We note that the data require ${f=0.09\pm0.03}$, which is broadly consistent with the combination of the calibration uncertainties reported by \citet{dob18} and the known typical flux density calibration accuracy of radio observations at the VLA (5\%). 

The new X-ray data independently support the evidence for temporally decaying  emission from GW170817. Comparing to the previous epoch of Chandra observations at $\approx 160$ days, and applying a simple binomial test, we find a probability of $P\sim 2\times 10^{-6}$ that the observed decrease in count-rate arises from a random statistical fluctuation.  We can thus reject the hypothesis of a random fluctuation  with  $\sim4.8\sigma$ confidence, and conclude that these Chandra observations provide the first statistically significant evidence for fading X-ray emission from GW170817.\footnotemark\footnotetext[3]{Repeating our MCMC analysis for the X-ray light curve alone, we find that the X-ray emission is consistent with a smoothed broken power law with $\alpha_2 < 0$ at $4.4\sigma$.}

Notably, the X-ray spectral index $\beta_x\equiv 1-\Gamma=-0.51^{+0.26}_{-0.27}$ is consistent with previously reported values (e.g., $\beta_x = -0.61 \pm 0.17 $ at $\sim160$ days; \citealt{Haggard17,marg17,marg18,Troja17,Troja18,Ruan18}). We observe no evidence for a steepening of the X-ray spectral index from $\beta_x$ to $\beta_x+0.5$, predicted when the synchrotron cooling frequency passes below the X-ray band {\citep{gs02}}. This behavior is consistent with our {successful} structured jet models, which predict that the cooling frequency will remain above the X-ray band until $\gtrsim10^4$ days \citep{marg18}, but rules out some dynamical ejecta models, which predict an earlier passage of the cooling break (e.g., \citealt{hot18,np18}). The radio-to-X-ray spectral index is $\beta_{\rm XR}=-0.583 \pm 0.013$ at $t\sim260$ days, confirming that the radio and X-ray bands still lie on the same spectral segment. This is fully consistent with the radio-to-X-ray spectral index of $\beta_{\rm XR}=-0.584\pm0.006$ reported at $t\sim160$ days \citep{marg18}, indicating no spectral evolution across the peak.  The radio-only spectral index at 217 days is $\beta_R={-0.74\pm0.20}$, {consistent with this value}. Finally, our {\it HST} observations are also consistent with the observed X-ray and radio decline.

\section{{Summary}}

We present new X-ray, optical, and radio observations of GW170817 at $\approx 220-{290}$ days. Our new broadband measurements show that the synchrotron emission from GW170817 has passed its peak brightness and has begun to decline. We find that the data continue to be well described by a single power law extending over $\approx 10^9-10^{18}$ Hz, with no sign of the spectral index change in the X-rays that would signify the passage of the synchrotron cooling break.  We find clear evidence ($\gtrsim 5\sigma$) for a turnover in the radio and X-ray light curves, and full agreement with the predicted evolution of our previously published {successful} structured jet models \citep{marg18,xie18}. The steep decline rate seen in our new radio and X-ray data is characteristic of all published {successful} structured jet models {\citep{laz17,xie18}}, and {mildly} disfavors some dynamical ejecta and pure cocoon models (e.g., {\citealt{hot18};} \citealt{np18,Troja18}). Continued observations will allow us to better constrain the post-peak decline rate, providing further insights into the ejecta structure.  If the emission continues to decay as predicted by the {successful} structured jet models, GW170817 should remain detectable with current radio and X-ray facilities until $\sim 1000$ days post-merger.

\acknowledgments
{We would like to thank the referee for helpful comments that improved this manuscript.} The Berger Time-Domain Group at Harvard is supported in part by the NSF through grant AST-1714498, and by NASA through grants NNX15AE50G and NNX16AC22G.
This work is partially based on observations acquired by the 
Chandra X-ray Observatory. The  Chandra X-ray Observatory Center is operated by the Smithsonian Astrophysical Observatory for and on behalf of the National Aeronautics Space Administration under contract NAS8-03060.
W.~F. acknowledges support from Program number HST-HF2-51390.001-A, provided by NASA through a grant from the Space Telescope Science Institute, which is operated by the Association of Universities for Research in Astronomy, Incorporated, under NASA contract NAS5-26555. B.~D.~M. is supported in part by NASA ATP grant NNX16AB30G.   The National Radio Astronomy Observatory is a facility of the National Science Foundation operated under cooperative agreement by Associated Universities, Inc.

\facilities{VLA, HST, CXO} 
\software{CASA, emcee, Numpy, pwkit}

\bibliographystyle{yahapj}
\bibliography{ms}

\end{document}